\documentclass[11pt]{article}
\def\be{\[}
\def\ben{\begin{equation}}
\def\ba{\begin{array}}
\def\bea{\begin{eqnarray}}
\def\ee{\]}
\def\een{\end{equation}}
\def\eea{\end{eqnarray}}
\def\ea{\end{array}}
\def\btab{\begin{table}}
\def\btabu{\begin{tabular}}
\def\etab{\end{table}}
\def\etabu{\end{tabular}}
\def\bit{\begin{itemize}}
\def\eit{\end{itemize}}

\def\la{\langle}
\def\ra{\rangle}
\def\pd{\partial}

\def\gb{\beta}

\def\e{\epsilon}

\def\om{\omega}

\def\l{\lambda}

\def\s{\sigma}

\def\Z{{\cal Z}}
\def\half{{\textstyle{1 \over 2}}}

\def\b1{{\bf 1}}

\def\br{\mbox{\boldmath $r$}}
\def\bj{\mbox{\boldmath $j$}}

\def\bQ{\mbox{\boldmath $Q$}}

\def\cosh{\hbox{cosh}}
\def\sign{\hbox{sign}}

\def\nn{\nonumber}
\def\bb{\left(}
\def\eb{\right)}
\def\bs{{\bf s}}
\def\br{{\bf r}}
\newcommand{\name}{\arabic{section}}
\newcommand{\newsection}[1]{\section{#1}\renewcommand{\theequation}
                              {\name.\arabic{equation}}
                            \setcounter{equation}{0}}

\def\bs{{\bf s}}

\usepackage{epsfig,latexsym}

\begin{document}
\title{Gauged $O(n)$ spin models in one dimension}
\author{{\bf M. Hasenbusch}\\
        \\
         Humbolt Universit\"{a}dt zu Berlin,\\
         Institut f\"{u}r Physik, Invaliden Str. 110, \\
         D10099 Berlin, Germany \\
        \\
        and\\
        \\
        {\bf  R.R. Horgan}\\
        \\
        D.A.M.T.P.\\ 
        Silver Street, Cambridge, England CB3 9EW}
\maketitle
\begin{abstract}
We consider a gauged $O(n)$ spin model, $n \ge 2~$, in one dimension which contains both the
pure $O(n)$ and $RP^{n-1}$ spin models and which interpolates between them. Various
formulations of this theory are given, one of which shows that it belongs to a general 
set of models for which it has been suggested that, for $n>2$, three distinct universality classes
exist and that $RP^{n-1}$ and $O(n)$ belong to different classes. We show that
our model is equivalent to the non-interacting sum of an $O(n)$ model and the Ising model
which allows a simple derivation of the complete mass spectrum that scales in
the continuum limit. We demonstrate that there are only two universality classes, one of
which contains the $O(n)$ and $RP^{n-1}$ models and the other which has a tunable parameter 
but which is degenerate in the sense that it arises from the direct sum of the $O(n)$ 
and Ising models.
\end{abstract}
\vfill
DAMTP-1998-158\\
 HUB-EP-98/74
\newpage
\newsection{\label{INTR}\bf Introduction}
Recently, there has been a considerable amount of discussion in the literature
\cite{soea:1,soea:2,hase,niea,scea,cuea,caea,seyi} concerning the nature 
of the universality classes of models which contain
the $O(n)$ spin model and the related $RP^{n-1}$ model. In particular,  
one-dimensional versions of these theories have been discussed \cite{cuea,caea,seyi}
in order to shed light on the mechanism behind the classification of the universality 
classes and to give an explicit calculation which demonstrate the results. 
For example, Caracciolo et.~al \cite{soea:1,soea:2} have given numerical evidence to support the
statement that in 2D the $RP^2$ and $O(3)$ spin models have different continuum 
limits which contradicts the usual belief that the universality classes of these
models are the same. In contrast, Niedermayer et.~al \cite{niea} and Hasenbusch \cite{hase}
argue that in 2D the $O(3)$ and $RP^2$ models do indeed belong to the same universality class
as long as one is careful to consider the right operators. In support of this
conclusion Catterall et.~al \cite{scea} using the Monte-Carlo renormalization
group conclude that the scaling observed numerically in $RP^2$ and assumed to
be associated with a continuum limit is, in fact, only apparent and that the $RP^2$
model is in the basin of attraction of the $O(3)$ fixed point although very large
correlation lengths would be needed to verify this fact. 

In 1D Campostrini et.~al \cite{caea}, Cucchieri et.~al \cite{cuea} and Seiler and Yildirim 
\cite{seyi} investigate the transfer matrix of general models with $O(n)$ spins
and they conclude that, for $n > 2$, there exist three distinct universality classes and that
$O(n)$ and $RP^{n-1}$ are not in the same class. This result is cited as support for the
similar conclusions in the case of 2D models. In this paper we consider a gauged $O(n)$ 
model in 1D, $n \ge 2$, similar to the 2D one discussed in \cite{scea} and we show that when the
gauge field in summed over the resulting action falls in the class of models investigated
by the authors cited above. We show that a field redefinition decomposes our model into
the non-interacting sum of an $O(n)$ spin model and an Ising model and that by considering
an appropriate correlation function the mass ratios $R_l$ computed by Seiler and Yildirim 
\cite{seyi}, which scale in the continuum limit, follow immediately from this observation. 
However, it is also evident from the Ising formulation that there are other scaling masses in 
the theory which are missed in their analysis because they do not directly correspond
to eigenstates of the transfer matrix in their formulation. These correspond to
the states which contribute to a different correlator in which local operators are
connected by a gauge string. In the Ising formulation the transfer matrix is well defined
and its eigenstates give rise to the full set of scaling masses characterizing the continuum
limit. In any of the formulations susceptibilities can be defined which are sensitive to
the complete set of states and which therefore define the continuum limit. The outcome is
that there are two universality classes one of which contains the $O(n)$ and $RP^{n-1}$ 
models and the other which has a tunable parameter. This second class is seen to arise
simply from the non-interacting sum of the $O(n)$ and Ising models and so is degenerate
in that it is an obvious way of producing an effect of this kind. The introduction
of Ising spins is closely related to the approach adopted by Niedermayer et.~al \cite{niea}

In section \ref{MODEL} we describe the gauge model and its various different manifestations;
in section \ref{SPECTRUM} we compute the spectrum; in section \ref{CONT_LIMIT} we calculate 
the scaling mass ratios for different choices of continuum limit, and in section \ref{CONCLUSION} 
we draw our conclusions.

\newsection{\label{MODEL}\bf The Model}
The action is given by
\ben
S(\bs,\s,\bj,\l)~=~\sum_x~\gb f\bb\bs_x\cdot\bs_{x+1}\eb\s_{x,x+1}~+~
                   \bj_x\cdot\bs_x~+~\l_{x,x+1}\,\s_{x,x+1}~,
                                                       \label{ACTION_G}
\een
where $\bs_x$ is a unit $n$-component vector spin field, $\s_{x,x+1}$ is a $Z_2$ 
gauge field, and $f\bb z\eb$ is an odd function that has its maximum in $[-1,1]$ 
at $z=1$ and the derivative $f^\prime(1) > 0$ exists. The fields $\bj_x$ and $\l_{x,x+1}$ 
are sources coupled to the spin and gauge fields respectively. This action is invariant 
under the gauge transformation
\bea
\bs_x~&\rightarrow&~h_x\,\bs_x~,\nn\\ 
\s_{x,x+1}~&\rightarrow&~h_x\,\s_{x,x+1}\,h_{x+1}~,\nn\\ 
\bj_x~&\rightarrow&~h_x\,\bj_x~,\nn\\ 
\l_{x,x+1}~&\rightarrow&~h_x\,\l_{x,x+1}\,h_{x+1}~, \label{GTR}
\eea
where the gauge transformation $h_x$ takes values in $Z_2~$. The partition 
function is 
\ben
\Z~=~\sum_{\{\s\} = \pm 1}\int d\{\bs\}~\exp\bb\,S(\bs,\s,\bj,\l)\,\eb~. \label{PFUNC}
\een
Here, $\l_{x,x+1}$ can be interpreted as a chemical potential for kinks, the 1D equivalent
of vortices. The model possesses pure $RP^{n-1}$ symmetry when $\bj_x = \l_{x,x+1} = 0,~\forall x$.
In the limit $\bj_x = 0,~\l_{x,x+1} \rightarrow \infty~\forall x$ the usual $O(n)$ model is 
recovered.  Thus varying  $\l_{x,x+1}$ allows us to interpolate between these two models and 
to study the small coupling region of these and the resulting hybrid theories.

In the general theory an important correlator to consider is 
\ben
G(x,y)=~\la\bQ(\bs_x)\cdot\prod_{r=x}^{y-1}\s_{r,r+1}\cdot\bQ(\bs_y)\ra~\left|_{~\bj=\l=0}\right.~,
\label{GRP2_G}
\een
where $\bQ(\bs)$ is an odd--parity tensor function of its argument: $\bQ(-\bs) = -\bQ(\bs)$~.
The form of this correlator is dictated by the requirement that for $RP^{n-1}$ it must be gauge 
invariant. Clearly, 
\ben
G(x,y)~=~\bQ\bb{\pd\over\pd\bj_x}\eb\cdot{\pd\over\pd\l_{x,x+1}}\ldots
         {\pd\over\pd\l_{y-1,y}}\cdot\bQ\bb{\pd\over\pd\bj_y}\eb\,\log{\cal Z}\left|_{\bj=\l=0}\right.~.
\label{G_DIFF}
\een
If we do the sum over the $\{\s\}$ gauge fields explicitly
in equation (\ref{ACTION_G}) we find the action as a function of the $\{\bs\}$ only to be
\ben
A(\bs,\bj,\l)~=~\sum_x~\gb\;g\bb\bs_x\cdot\bs_{x+1},\l_{x,x+1}\eb~+~\bj_x\cdot\bs_x~,
\label{ACTION_S}
\een
with
\ben
g(z,\l)~=~\log\left[\;\cosh(\gb\,f(z)~+~\l)\;\right]~. \label{FNG}
\een
For $\l_{x,x+1} = 0,~\forall x$ we can, instead, consider the $RP^{n-1}$ invariant action
$g(z,0)$ as given and then $f$ is determined by the inverse of this procedure:
\ben
f(z)~=~{1\over\gb}\,\sign(z)\,\cosh^{-1}\left[\exp(\,\gb g(z,0)\,\right]~,
\een
where the positive branch of $\cosh^{-1}$ is chosen. Using equation(\ref{G_DIFF}) we find
\ben
G(x,y)~=~\la\bQ(\bs_x)\cdot\prod_{r=x}^{y-1}\om_{r,r+1}\cdot\bQ(\bs_y)\ra~\left|_{~\bj=\l=0}\right.~,
\label{GRP2_S}
\een
where 
\ben
\om_{x,x+1} = \tanh\left[\;\gb f\bb\bs_x\cdot\bs_{x+1}\eb\;\right]~.\label{OM}
\een
The composite fields $\{\om\}$ play the r\^ole of gauge-like fields. We are interested in the 
physics near to the transition point at zero temperature, i.e., $\gb \rightarrow \infty~$.  
In this limit the configurations that contribute appreciably are those for which 
$\gb\left|f\right| \gg 1$ and so we have
\ben
\om_{x,x+1}~=~\tanh(\gb f(\bs_x\cdot\bs_{x+1}))~\stackrel{\gb \rightarrow \infty}{\longrightarrow}
~\sign(\bs_x\cdot\bs_{x+1})~.
\label{OM_BIG}
\een
For example, for $\gb \rightarrow \infty$ the action $A(\bs,\bj,\l)$ for pure $RP^{n-1}$ 
behaves like
\ben
A(\bs,\bj,\l)~\sim~\sum_x\;\gb\left|f(\bs_x\cdot\bs_{x+1})\right|~. \label{A_NOGAUGE}
\een
The two formulations for the action in equations (\ref{ACTION_G}) and (\ref{ACTION_S}) are
entirely equivalent with correspondingly equivalent correlators defined in equations
(\ref{GRP2_G}) and (\ref{GRP2_S}). In the latter case the r\^ole of the gauge field is
played by the link variable $\{\om\}$ defined in equations (\ref{OM}) and (\ref{OM_BIG}).

For the case of $O(n)$ the eigenfunctions of the transfer matrix are the harmonic functions $Y_l(\bs)$ 
on $S_{n-1}$, and in the special case where $f(z) = z$ the model is the discrete approximation to the quantum 
mechanics of a particle constrained to $S_{n-1}$ which has eigenfunctions $Y_l(\bs)$ and associated energy 
eigenvalues $E_l = l(l+n-2)/4\gb$. 

For the general model the transfer matrix can be written as ($\bj_x = 0, \forall~x$)
\bea
\lefteqn{T(x,x+1;\bs,\s,\l) = }\nn\\ 
&&\exp\;(\l_{x,x+1}\s_{x,x+1})\left[ 1+\sum_{l~\mbox{even}}\mu_l(\gb)\;Y_l(\bs_x)Y_l(\bs_{x+1}) \right.
                          +\left.\sum_{l~\mbox{odd}}\nu_l(\gb)\;\s_{x,x+1}\;Y_l(\bs_x)Y_l(\bs_{x+1}) \right]~. \nn\\
&&                                                                                                 \label{TR}
\eea
The general set of correlators takes the form 
\bea
F_l(x,y)&=&\la Y_l(\bs_x)\; Y_l(\bs_y) \ra~, \label{FL} \\ 
G_l(x,y)&=&\la Y_l(\bs_x)\;\prod_{r=x}^{y-1}\s_{r,r+1}\;Y_l(\bs_y) \ra~. \label{GL}
\eea
For the $O(n)$ model $F_l = G_l$, and in the $RP^{n-1}$  case $G_l = 0$ for $l$ even, and $F_l = 0$ for $l$ odd whenever
$x \ne y$. 

In formulation without explicit gauge field, such as is given in equation (\ref{A_NOGAUGE}), the transfer matrix 
is given (up to an overall irrelevant normalization) by
\bea
\lefteqn{T^\prime(x,x+1;\bs) = } \nn\\
&&  1~+~\sum_{l~\mbox{even}}\mu_l(\gb)\;Y_l(\bs_x)Y_l(\bs_{x+1})
     ~+~\tanh(\l_{x,x+1})\;\sum_{l~\mbox{odd}}\nu_l(\gb)\;Y_l(\bs_x)Y_l(\bs_{x+1})~,\nn\\
&&                                                                               \label{TR_NOGAUGE}
\eea
and then the correlator $G_l$ takes the form
\ben
G_l(x,y) = \la Y_l(\bs_x)\;\prod_{r=x}^{y-1}\om_{r,r+1}\;Y_l(\bs_y) \ra~, \label{GL_NOGAUGE}
\een
whereas the expression for $F_l$ is unchanged.

The question to be addressed is whether the $G_l$ correlators correspond to the correlation of local 
operators which interpolate states each labelled by its energy. In the next section we shall show
that this is the case and we shall compute the mass ratios which characterize the continuum limit,
$\gb \rightarrow \infty$.

\newsection{\label{SPECTRUM}The spectrum}

In the previous section we have shown the equivalence between two formulations of the 1D theory which 
includes both the $O(n)$ and $RP^{n-1}$ models and which interpolates between them. The equivalence 
between the respective formulations of the relevant correlation functions was also demonstrated. Hence,
it is sufficient to determine the spectrum of the model using the version formulated with an explicit
gauge field.

We first perform a field redefinition. Define $\br_x$ and $\e_x$ by
\ben
\s_{x,x+1} = \e_x\e_{x+1}~,~~~~~~~~\br_x = \e_x\bs_x~. \label{FIELD_REDEF}
\een
This is possible because the field $\s_{x,x+1}$ can be always written as a pure gauge in one dimension.
We also note that the definition of $\br_x$ corresponds to the general transformation
\be
\bs_x = \prod_{z=-\infty}^{x-1}\s_{z,z+1}\;\br_x~.
\ee
Since $f(z)$ is an odd function the action now becomes 
\ben
S(\bs,\s,\bj,\l)~=~\sum_x~\gb f\bb\br_x\cdot\br_{x+1}\eb~+~
                   \e_x\bj_x\cdot\br_x~+~\l_{x,x+1}\,\e_x\e_{x+1}~.
                                                       \label{ACTION_OI}
\een
The model has decoupled into an $O(n)$ spin model plus an Ising model with inter-site
coupling $\l_{x,x+1}$. The correlators (\ref{FL}) and (\ref{GL}) can now be expressed as
\ben
\ba {lcl}
F_l(x,y)&~~~~=~~~~&\left\{
\ba {ll}
\la Y_l(\br_x)\;Y_l(\br_y) \ra_O~,&l~~\mbox{even} \\
\la Y_l(\br_x)\;Y_l(\br_y) \ra_O\;\la \e_x\;\e_y\ra_I~,~~~~&l~~\mbox{odd} 
\ea
\right.\\
\\
G_l(x,y)&=&\left\{
\ba {ll}
\la Y_l(\br_x)\;Y_l(\br_y) \ra_O\;\la \e_x\;\e_y\ra_I~,~~~~&l~~\mbox{even} \\
\la Y_l(\br_x)\;Y_l(\br_y) \ra_O~,&l~~\mbox{odd} 
\ea
\right.

\ea
\label{FLGL_OI}
\een
Where the subscripts $O$ and $I$ signify a correlator in the pure $O(n)$ model and the
decoupled Ising model, respectively. 

We consider the situation where $\bj_x=0, \forall x$ and $\l_{x,x+1} = \mbox{constant} \equiv \l$. 
There is a residual global symmetry of $\e \rightarrow -\e,~\forall x$ which is not in the
original model. This symmetry can be accounted for by fixing $\e_{-\infty} = 1$ but is unimportant
for $\l < \infty$ since the Ising system is disordered. At the transition point, $\l = \infty$, this
condition fixes the magnetization, and in this particular case the Ising spins are frozen with 
$\la \e_x\;\e_y\ra_I = 1$ and so $F_l$ and $G_l$ are not distinct. There is one set of 
correlators corresponding to the $O(n)$ model.

As $\gb \rightarrow \infty$ a continuum limit is approached and the dominant fluctuations in $\bs$ 
are controlled by $f(z)$ near to $|z|=1$. So, since $f(z)$ is an odd function, the important contributions 
to the partition function can be calculated for the model in which we make the replacement
\ben
f(z) \rightarrow f^\prime(1)z~.
\een
This factor $ f^\prime(1)$ can be absorbed into redefinition of $\gb$: $\gb^\prime =  f^\prime(1)\gb$~,
and so in the continuum limit, from equation (\ref{ACTION_OI}) and without loss of generality, 
the spectrum of the $O(n)$ part of the model coincides with the spectrum of the quantum mechanics 
of a particle constrained to $S_{n-1}$ as was discussed as a special case in the previous section. 
Since we are interested in mass ratios as the continuum is approached we confine the discussion to 
the model where $f(z)=z$ and drop the prime on $\gb$, although we emphasize that this is not a special 
case but applies to all actions of the form given in equation (\ref{ACTION_G}) as the continuum 
limit is approached.

The interpolating states have wavefunctions taking values on the direct product space $S_{n-1} \otimes Z_2$, 
where the $Z_2$ is the space associated with the Ising degrees of freedom. The full spectrum of masses is
therefore 
\bea
m_l&=&l(l+n-2)/4\gb~, \nn\\
M_l&=&{l(l+n-2)~+~4c(\l)\over 4\gb}~, \label{MASSES}
\eea
where we define $c(\l) = \gb\log(\coth(\l)) \Rightarrow \l = \half\log(\coth(c/2))$. The mass 
spectrum of the interpolating states for $F_l~(G_l)$, which we denote $\mu^F_l~(\mu^G_l)$, is 
then $m_l$ for $l$ even (odd) and $M_l$ for $l$ odd (even).  

For states to survive in the continuum limit their associated masses must vanish in this limit: all
other states are lattice artifacts. For both $m_l$ and $M_l$ this will only happen as $\gb \rightarrow \infty$, 
as expected, but for $M_l$ we also require that in this limit $\l$ behaves such that $c(\l)$ is finite or 
diverges with $\gb$ more slowly than O($\gb$). It is the behaviour of $c(\l)$, and hence of $\l$, that 
distinguishes between the different kinds of continuum theory obtained in the limit $\gb \rightarrow \infty$,
and this will lead to a classification of the universality classes that are possible. 

\newsection{\label{CONT_LIMIT}The continuum limit}
To probe the continuum limit we can use the spectrum obtained from $F_l$ and $G_l$ above. In \cite{seyi} 
the authors discuss the transfer matrix of models which include the models considered here when they have 
been recast into the form given in equations (\ref{ACTION_S}) and (\ref{FNG}). Whilst there is not a one-to-one 
correspondence we shall show that our models give results which correspond to each of the universality classes  
cited in \cite{seyi} and so, by universality, our results are pertinent to the gamut of models considered in 
this reference. From the discussion in section \ref{MODEL} and equations (\ref{FL}) and (\ref{GL_NOGAUGE}) we see
that to access all possible states in the case where the gauge field has been summed over we must 
introduce the ``gauge string'' formed from the $\om_{x,x+1}$. In 1D this will still give rise to
local interpolating operators. To just consider the transfer matrix and omit these degrees of freedom
corresponds to just discussing the spectrum of states contributing to $F_l$ but not $G_l$. This is the
case for the analysis in \cite{seyi}. To facilitate the first part of the discussion we take $\l$ to 
depend on $\gb$ so that as $\gb \rightarrow \infty$
\ben
\ba{llll}
&c &\sim& c_0 \gb^\eta~, \\ 
\Longrightarrow~~~~~~~~~~&&& \\
&\l(\gb)~&\sim&\left\{ 
\ba{ll}
\half\bb(1-\eta)\log(\gb) + \log(2/c_0)\eb~,~~~~~~&\eta < 1 \\
\half\log(\coth(c_0/2)~,&\eta = 1 \\
\exp(-c_0\gb^{\eta-1})~,&\eta > 1
\ea\right.
\ea
\een
where $c_0$ is a constant. This is not the most general parametrization but it suffices to 
elucidate the important cases and the conclusions will not be changed by using a more general form. 
For $\eta < 1$ the Ising states survive into the continuum limit, but for $\eta \ge 1$ they are lattice
artifacts only and are not relevant as $\gb \rightarrow \infty$ and so we must recover the spectrum of
the $O(n)$ model in the limit. The dependence of the behaviour of $\l(\gb)$ on $\eta$ is significant and
takes the form 
\bea
\eta < 1~~~&&\lim_{\gb \rightarrow \infty} \l(\gb) = \infty~,\nn\\
\eta = 1~~~&&\lim_{\gb \rightarrow \infty} \l(\gb) = \half\log(2/c_0)~,\nn\\
\eta > 1~~~&&\lim_{\gb \rightarrow \infty} \l(\gb) = 0~.  \label{LLIMIT}
\eea
Consequently, for $\l = 0$ we must recover the spectrum of the $O(n)$ model in the continuum
limit. This shows that the $RP^{n-1}$ model, for which $\l = 0$, has the same continuum
limit as the $O(n)$ model which has $\l = \infty$. We shall further elucidate this result below. 

We consider the continuum limit mass ratios $R_l = \mu^F_l/\mu^F_1$ and identify the different classes:

\ben
R_l = \left\{
\ba{llll}
{l(l+n-2) \over n-1}~,&\eta < 0~,&c \rightarrow 0~~~~&{\bf I}\\
\\
{l(l+n-2)\;+\;4c_0\;(1-(-1)^l)/2 \over (n-1) + 4c_0}~,~~~~~~&\eta = 0~,&c \rightarrow c_0&{\bf II}\\
\\
{1-(-1)^l \over 2} ~.&0 < \eta < 1~,&c \rightarrow \infty&{\bf III}
\ea
\right. \label{RL}
\een
These mass ratios are identical to those quoted for classes {\bf I,~II,~III} in reference \cite{seyi}.

We could similarly discuss the ratios $\mu^G_l/\mu^G_1$ associated with correlators $G_l$ but it
is better to consider the complete spectrum $\{m_l,M_l\}$ given by $F_l$ and $G_l~$ and define the mass 
ratios to be $P^{(F)}_l = \mu^F_l/m_1$ and $P^{(G)}_l = \mu^G_l/m_1$. The nature of the continuum limit is determined
by considering the $P_l$ rather than the $R_l$ which are incomplete. We then find the three classes above 
to be characterized by

\ben
P^{(F/G)}_l = \left\{
\ba{llll}
{l(l+n-2) \over (n-1)}~,&\eta < 0~,&c \rightarrow 0~~~~&{\bf I}\\
\\
{l(l+n-2)\;+\;4c_0\;(1\mp(-1)^l)/2 \over (n-1)}~,~~~~~~&\eta = 0~,&c \rightarrow c_0&{\bf II}\\
\\
{l(l+n-2) \over (n-1)}~~\bb{l~\mbox{odd} \atop l~\mbox{even}}\eb~,~~\infty~~\bb{l~\mbox{even} \atop l~\mbox{odd}}\eb~,&
0 < \eta < 1~,&c \rightarrow \infty&{\bf III}
\ea
\right. \label{PL}
\een
A ratio of $\infty$ means that the corresponding state is not in the spectrum. 

The nature of the three classes is now clear. Classes {\bf I} and {\bf III} have the same values for the $P_l$:
in {\bf I} the Ising spins are frozen, as remarked earlier, and in {\bf III} the Ising mass gap scales but with
infinite ratio to the lowest lying states. In both cases we recover the universality class of the pure $O(n)$ spin 
model. For class {\bf II} the $P_l$ are labelled by the continuous parameter $c_0$ which Seiler and Yildirim take 
to label the continuum of universality classes. 

It remains to discuss the cases with $\eta \ge 1$. In these cases the mass $\mu^F_1$ does not survive into the 
continuum and so the ratios $R_l$ have no meaning. From equation (\ref{LLIMIT}) we have that $0 \le \l < \infty$ and
the result for the $P_l$ falls into class {\bf III} of equation (\ref{PL}). We conclude that all models
with finite non-negative $\l$ belong to the $O(n)$ universality class as was stated earlier. In particular,
the $RP^{n-1}$ model belongs to the $O(n)$ universality class. This result is trivially extended to cover
all $\l$, $-\infty \le \l < \infty$.

The universality class {\bf I} and {\bf II} are associated with the fixed point at $(\gb = \infty, \l = \infty)$,
and the parameter $c_0$ controls the way in which the trajectories of {\bf II} approach this point.  

\section{\label{CONCLUSION}Conclusions}

We have considered a class of gauged $O(n)$ models in one dimension, equation (\ref{ACTION_G}), which includes 
$RP^{n-1}$. The gauge field can be summed over and the resultant alternative form for the action, equation
(\ref{ACTION_S}), falls in the class of actions considered by Seiler and Yildirim \cite{seyi}, Campostrini et.~al
\cite{caea} and Cucchieri et.~al \cite{cuea}. The conclusion of these cited works concerning such actions is 
that there exist three distinct universality classes and that, in particular, $RP^{n-1}$ is not in the same class as $O(n)$.
In this paper we have shown that gauged $O(n)$ version of the model can be analyzed in a very straightforward 
manner and that the scaling mass ratios, $R_l$, quoted by Seiler and Yildirim can be clearly and simply derived, 
equation (\ref{RL}), and their origin elucidated to be from the non-interacting sum of an $O(n)$ and Ising model.  
Cucchieri et.~al  note briefly that Ising degrees of freedom are present in these models but do not comment further 
on their remark. 

The analysis of Seiler and Yildirim, Campostrini et.~al, and Cucchieri et.~al is based on the transfer matrix formulation 
of the theories which coincides with the analysis of the correlators $F_l$ alone, equation (\ref{FL}), and the masses of the
corresponding interpolating states. However, the universality class of a model is described by the properties of the 
model as the critical point is approached and, in particular, how various susceptibilities diverge. The susceptibilities
are defined in terms of the derivatives of the free energy with respect to external fields and it is clear from
either version of the model we have studied that the correlators $G_l$ legitimately define a susceptibility through
equations like equation (\ref{G_DIFF}). However, the form for $G_l$ given in (\ref{GRP2_S}) shows that the masses
of the interpolating states will not be determined by an analysis of the transfer matrix (\ref{TR_NOGAUGE}) relevant
to the model after the gauge field has been summed over. For the gauge model the transfer matrix in the Ising formulation
is simply
\ben
T_I(x,x+1,\bs)~=~\exp\:(\gb\;\br_x\cdot\br_{x+1})\left(
\ba{ll}
\exp(\l_{x,x+1})~~~~&\exp(-\l_{x,x+1})\\  
\\
\exp(-\l_{x,x+1})~~~~&\exp(\l_{x,x+1})\\ 
\ea
\right)~, 
\label{TR_ISING}
\een
and the state basis is then spanned by
\ben
Y_l(\bs) \left(\ba{r}1\\1\\\ea\right)~,~~~~~~Y_l(\bs) \left(\ba{r}1\\-1\\\ea\right)~.
\label{BASIS}
\een

In the non-gauge formulation it is not obvious that the $G_l(x,y)$ decay exponentially as $|x-y| \rightarrow \infty$
but we have shown that they do possess this property  and that the masses of the corresponding states scale into 
the continuum limit. When the complete mass spectrum is taken into account we obtain the scaling mass ratios $P_l$ 
in equation (\ref{PL}) which show that the universality classes {\bf I} and {\bf III} are the same. In particular, 
$RP^{n-1}$ belongs to the same universality class as $O(n)$. The existence of a continuous set of universality 
classes of type {\bf II} is seen to be degenerate in the sense that it arises from the non-interacting sum of two 
independent models -- a construction which is always available to produce such a result. 

The results of this paper confirm in one-dimension the results obtained by Niedermayer et.~al \cite{niea} and
Hasenbusch \cite{hase} in their study of similar models in two dimensions. Niedermayer et.~al also introduce
an Ising degree of freedom to reach their conclusions.

\bibliography{refs}
\bibliographystyle{unsrt}

\end{document}